**Influence of dislocation density on the tribological response in oxides: case study on $SrTiO_3$**

Chukwudalu Okafor[1], Oliver Preuß[1]*, Thomas Chudoba[2]*, Ujjval Bansal[1], Daniela Exner[1,4], Konrad Priszokovich[1], Yanfei Gao[3], Christoph Kirchlechner[1], Xufei Fang[1]

[1]Institute for Applied Materials, Karlsruhe Institute of Technology, 76131 Karlsruhe, Germany

[2]ASMEC GmbH, 01190 Dresden, Germany

[3]Department of Materials Science and Engineering, University of Tennessee, 37996 Knoxville, TN, USA

[4]Karlsruhe Nano Micro Facility (KNMFi), Karlsruhe Institute of Technology, 76344 Eggenstein-Leopoldshafen, Germany

*Corresponding authors: oliver.preuss@kit.edu (O.P); t.chudoba@asmec.de (T.C)

**Abstract**

Most ceramics suffer from brittle surface damage and cracking when small particles slide across the surface under low load. Microscratching tests are a useful technique to mimic this loading scenario while retaining the material's deformation history; however, the impact of pre-seeded dislocations, which can significantly facilitate plastic deformation, has not been explored in ceramics. Here, the influence of mechanically seeded dislocations on the microscratching response of oxides is investigated using a model perovskite, $SrTiO_3$. First, various dislocation densities over four orders of magnitude are introduced via room-temperature cyclic Brinell indenter scratching. Subsequently, load-ramped microscratching tests are performed within the pristine and dislocation-seeded regions using a nominally 30 µm spherical diamond tip. On the reference pristine surface, we observe a clear transition from elastic to elasto-plastic deformation, followed by median/radial cracking at higher loads. In contrast, pre-seeded dislocations, accompanied by residual compressive stresses, suppress elastic deformation and lead to median/radial crack shortening and subsequent transition to a partial cone crack. The subsurface cracks are characterized by 3D Nano-CT. The changes in crack geometry, with and without dislocations, were described using the Lawn-Evans-Marshall and Lawn-Wiederhorn-Roberts models. These findings provide direct evidence of dislocation-regulated near-surface damage tolerance, with general applicability to other plastically deformable oxides.

## 1. Introduction

Oxides are widely used in functional applications where mechanical reliability is often governed by near-surface damage. During processing, handling, polishing, grinding, or sliding contact, highly localized stresses can generate subsurface cracks, median/radial cracks, chipping, and plastic zones [1-3], even when the nominal bulk stresses remain relatively low. Such damage is particularly critical for functional oxides across a wide range of applications where surface finishing and resistance to cracking are vital for performance, including semiconducting [4, 5] and medical applications [6, 7], compromising both mechanical integrity and device performance [2].

Microscratching test provides a controlled way to mimic these contact-driven damage processes, while it allows for simultaneous tracking of the transitions from elastic deformation to the onset of plasticity and cracking under a continuously increasing load. This enables evaluation of damage initiation, including plastic flow, crack initiation, and crack propagation, which is beneficial for failure prediction. For more than 60 years, numerous attempts at understanding the mechanical response of ceramics under sliding contact [1, 8-10] have been documented. The earlier focus was on the friction and strength (resistance to failure) of ceramics (mostly rock salts) [8, 9]. In recent years, efforts have shifted to understand the cracking behavior, particularly during the post-processing of ceramic devices [3, 11-13]. In particular, Lawn and co-authors investigated various engineering-relevant oxides (such as alumina, $Al_2O_3$) and identified the threshold damage mechanism and impact on failure [3]. However, as the focus was on brittle ceramics, it has been widely perceived that there is a direct transition from elastic deformation to cracking, with less attention being paid to assessing the onset of dislocation plasticity and its impact on the observed cracking behavior.

Over the past two decades, there has been a renewed interest in dislocation-tuned functionalities in ceramics [14-16], which further sparked studies in dislocation mechanics spanning the length scales [17, 18]. For instance, dislocations in $SrTiO_3$ and other perovskite oxides (such as $KNbO_3$, $KTaO_3$) have been successfully introduced at the bulk scale at room temperature without cracking [19-22], and the dislocation density can be tuned over several orders of magnitude, from $10^{11}$ $m^{-2}$ up to $10^{15}$ $m^{-2}$ [23]. The most successful techniques involved in dislocation engineering in these oxides include cyclic Brinell indentation and cyclic Brinell indenter scratching [19, 20]. For both approaches, a hardened steel ball is employed to tune the dislocation density by increasing the number of indentation cycles or scratch passes. Mechanical indentation or scratching of hard ceramics using softer metal sliders is a proven method to introduce dislocations, previously established for ionic crystals with rock-salt structure, such as MgO [5, 6]. The current authors observed that intentional seeding of dislocations prior to point-loading, such as Vickers indentation, in e.g., MgO, $KNbO_3$ and $SrTiO_3$ resulted in consistent indentation crack shortening as dislocation density increased [23-25]. The resulting crack shortening and suppression is also partly attributed to the residual compressive stresses introduced during the mechanical seeding of dislocations [18, 23, 24].

Despite these improvements in plasticity and crack suppression, it remains an open question how controlled pre-engineered dislocations affect the microscratching response of functional oxides. As dislocations are the carriers of plastic deformation in crystalline solids, one of the bottlenecks leading to limited plasticity and, in turn, brittle fracture in typical ceramics is their low dislocation density. Suppose the dislocation density in the samples could be increased by several orders of magnitude before mechanical scratching; how different would the tribomechanical responses of ceramics be? Could readily available dislocations act as dislocation sources, leading to further plastic deformation and suppressing crack formation during the scratch test? Furthermore, suppose the applied load during microscratching is varied (e.g., load-ramp), and the pre-seeded dislocations do show significant crack suppression (as observed during Vickers indentation in $SrTiO_3$ [18]), would there be a threshold load beyond which the dislocations would fail to suppress cracking?

Here, a two-step scratching approach was adopted to address the open question. First, a Brinell indenter equipped with a hardened steel ball (tip diameter 2.5 mm) is used to generate dislocation-rich regions with controlled dislocation density over a large area at the mesoscale [20]. A subsequent load-ramped microscratching test was performed in the dislocation-rich and reference pristine regions using a much smaller diamond spherical indenter (nominal tip radius 30 µm) to probe the resulting deformation and cracking response. A combination of SEM and Nano-CT was employed to unveil the surface and bulk features of the deformation and cracking response. Finally, Lawn-Evans-Marshall (LEM) and Lawn-Wiederhorn-Roberts (LWR) crack models were adopted to describe the controlling mechanism for the cracking and crack transitions with dislocation density.

## 2. Experimental procedure

### 2.1. Material selection and sample preparation

A nominally undoped strontium titanate ($SrTiO_3$) single crystal with dimensions 10 mm x 10 mm x 1 mm and (001) surface orientation (Crystec GmbH, Berlin, Germany) was used for the scratching test. Here, $SrTiO_3$ is chosen due to its wide use as a benchmarking model system for studying room-temperature dislocation plasticity in ceramic materials, and the findings on dislocation plasticity have also been extended to other plastically deformable ceramics at room temperature [26]. The as-received samples were cut into four pieces (5 mm × 5 mm × 1 mm), one of which was chemically etched in a solution containing 15 mL of 50% $HNO_3$ and 16 drops of 50% HF for 20 seconds to determine the pre-existing dislocation density prior to mechanical deformation. This dislocation density is used as a representative value for the other pristine samples from the same batch.

### 2.2. Mechanically seeded dislocations via Brinell indenter scratching

To assess the impact of dislocation density on the response of single-crystal $SrTiO_3$ during microscratching, dislocations were first introduced via cyclic Brinell indenter scratching in the near-surface region of the pristine crystals, which generates a plastically deformed scratch track of hundreds of micrometers in width. Note that this Brinell indenter scratching step solely aims to introduce a scratch track with high dislocation densities, which scale with the number of passes as described in a previous work [20]. For this purpose, a universal hardness testing setup (Karl-Frank GmbH, Weinheim-Birkenau, Germany) equipped with a hardened steel ball (2.5 mm diameter, 1 kgf load) and an automated two-axis stage (Thorlabs Inc., Newton, NJ, United States) was used. Bidirectional cyclic scratching (except for the single-pass case, **Fig. 1A**) was performed to produce scratch tracks of a length of 4 mm and a width of 100 μm, with a scratching speed of 0.5 mm/s. The numbers of different passes 1x, 5x, and 20x, where *x* indicates the number of passes, are illustrated in **Fig. 1B**. The estimated dislocation densities for each scratching condition are: 1x ~ $10^{11}$ $m^{-2}$, 5x ~ $10^{13}$ $m^{-2}$, and 20x ~ $10^{14}$ $m^{-2}$ [20], while the pre-existing dislocation density in an unscratched, pristine sample is ~ $10^{10}$ $m^{-2}$.

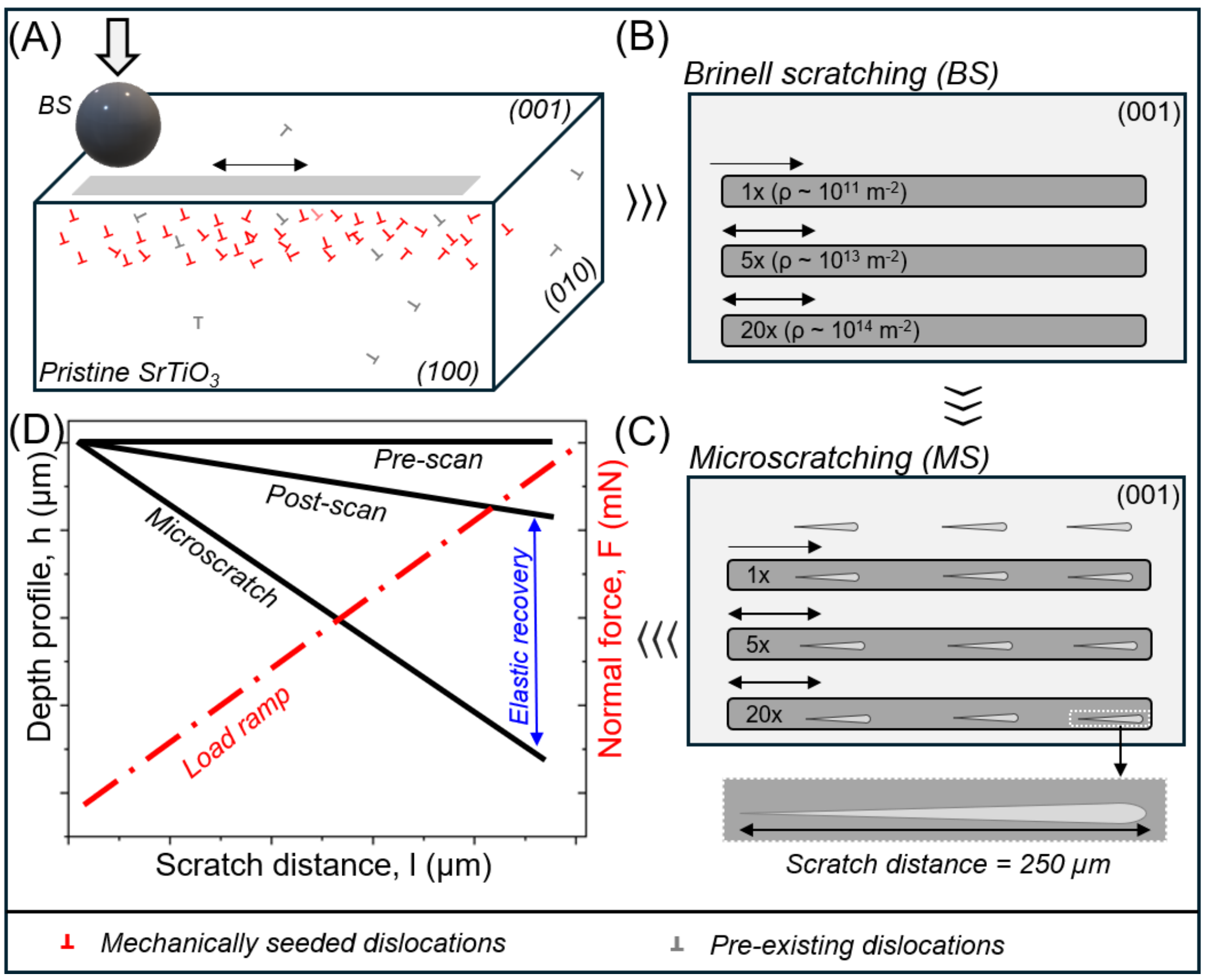

***Figure 1.*** *(Clockwise direction) Workflow for Brinell indenter scratching (BS) and microscratching (MS). (A): Pre-existing dislocations (grey) and mechanically seeded dislocations (red). (B): Brinell indenter scratching up to 20x (with various dislocation densities). (C): Microscratching tests in the pristine and dislocation-rich regions. (D): Illustration of the plotted data, pre-scan, scratch, and post-scan, and applied normal force (load ramping up to 600 mN). Note that* ***Fig. 1D*** *is for schematic illustration only: during the microscratching test, the normal load was ramped linearly with scratch distance; the corresponding penetration depth is expected to evolve nonlinearly because of the spherical contact geometry, elastic-plastic deformation, and increasing*

*contact area. The apparent linear trends in **Figs. 2** and **3** should therefore not be interpreted as a linear load-depth relation.*

### 2.3. Load-ramped microscratching test

After the dislocation-seeding process using cyclic Brinell indenter scratching, we investigated the tribological response of $SrTiO_3$ in the sample areas with low (~ $10^{10}$ $m^{-2}$), medium (~ $10^{11-13}$ $m^{-2}$), and high (~ $10^{14}$ $m^{-2}$) dislocation densities. To this end, a universal nanomechanical tester (ZHN-S, ASMEC GmbH, Dresden, Germany), equipped with a diamond spherical indenter (nominal tip radius 30 μm) was used to induce microscratching tracks of a length of 250 μm, with a maximum normal force of 600 mN (load-ramping starts from 0 mN) at a speed of 10 μm/s and a data acquisition rate of 16 Hz. For each Brinell indenter scratch track (including the pristine region), three representative microscratches were performed to ensure reproducibility (**Fig. 1C**). Prior to and after each microscratching test, the sample surface was scanned using the indenter tip itself and a scan force of 1 mN (termed as 'pre-scan' and 'post-scan', respectively, as schematically illustrated in **Fig. 1D**). The comparison allows observation of the topographical changes.

### 2.4. Surface morphology and crack characterization

A laser confocal microscope and a scanning electron microscope (SEM) were used to capture the dislocation distribution and cracking behavior (after chemical etching) induced by the scratching test. First, an overview of the microscratch tracks was captured using the laser confocal microscope (LEXT, OLS4000, Olympus IMS, Waltham, USA), which allows for a wide field of view (256 μm x 256 μm with the 50x objective). The surface topography of the Brinell indenter scratch tracks was also observed using the laser confocal microscope. Subsequently, to examine the generated dislocation etch pits and cracking behavior, samples were observed in the SEM (Merlin Gemini 2, Carl Zeiss Microscopy GmbH, Oberkochen, Germany). Secondary electron (SE) imaging mode (5.0 kV, working distance 10 mm) was used to capture the local regions of interest on the scratch tracks.

### 2.5 FIB milling and nano-computed tomography

SEM and laser microscope characterization are surface-limited techniques; therefore, nano-computed tomography (Nano-CT) was used to resolve the depth-dependent crack behavior. Prior to Nano-CT, site-specific micropillars containing representative median/radial (5x) and partial cone cracks (20x) were prepared. Coarse material removal was performed by femtosecond laser ablation (50 kHz pulse frequency, 400 mm/s, and 50% power) followed by a two-step Ga+ focused ion beam (FIB) milling. An initial high-current step (30 kV, 15 nA) reduced the diameter down to 25 μm, followed by a low-current finishing step (30 kV, 1.5 nA) to minimize redeposition. The free-standing cylindrical pillars with a final diameter and height of 20 μm were produced by this method. Note that the attenuation length of $SrTiO_3$ is ~12 μm at an X-ray photon energy of 5.4 keV; hence, the optimal micropillar diameter is ~1.5 times the attenuation length. This justifies the choice of 5x (median/radial crack) and 20x (partial cone crack) samples for the Nano-CT, since their crack surface lengths are less than 20 μm.

The micropillars were scanned using a lab-based X-ray microscope, Zeiss Xradia 810 Ultra (Nano-computed tomography, Nano-CT), equipped with a rotating Cr anode at 5.4 keV X-ray energy and a low-resolution field of view of 65 x 65 $μm^2$. The transmitted X-rays are recorded on a scintillator coupled to a CCD camera; the 20x magnification yields a FoV of 65 μm, a low-resolution limit of ~150 nm, and a voxel size of 0.126 μm. 901 projections were acquired over a 180° scan, with 0.2° rotation per scan, for ~35 h. For each crack, the crack depth was determined by tracking the crack through the reconstructed image stack from the top surface until the crack was no longer visible. Crack depth was calculated as the product of the number of slices and the voxel size (0.126 μm). This procedure was applied to all identified median/radial and partial cone cracks. The projections were 3D reconstructed using the proprietary Zeiss Scout and Scan Control System Reconstructor software. The reconstructed data were visualized using the ORS Dragonfly software (version 2022).

## 3. Results and Analyses

The two-step scratching approach (first Brinell indenter scratching at mesoscale, followed by microscratching at microscale) brings new insights into the tribomechanical response of single-crystal $SrTiO_3$ at room temperature. First, the observed behavior of pristine $SrTiO_3$ (without Brinell indenter scratching) is presented, focusing on two major transitions: the elastic-to-plastic and the plastic-to-cracking transitions. Second, the influence of mechanically seeded dislocations as well as their densities on the cracking behavior, namely, (i) crack suppression, and (ii) change in the crack geometries, was presented.

### 3.1. Microscratch response of pristine $SrTiO_3$

The microscratch response of pristine $SrTiO_3$ crystal was first examined to establish a benchmark before mechanical dislocation seeding. **Figure 2A** provides an overview of the microscratch, with the onset of plastic deformation highlighted by an arrowhead. First slip activity on the $\{110\}_{45}$ slip plane is identified by the dislocation etch pits captured by the laser microscope (**Figs. 2B-C**). Room-temperature active slip system in $SrTiO_3$ is $<1\bar{1}0>\{110\}$ [27]. Under a load-ramped microscratching test, the deformation response evolved progressively with increasing scratch distance and correspondingly increasing normal load. As illustrated in **Fig. 2C-D**, with an increasing load up to a scratch distance of ~26 μm (73 mN), only recoverable elastic deformation was detected by comparing the pre- and post-scan profiles. This agrees with the observation of dislocation etch pits, which appeared at a scratching distance of ~26 μm (73 mN) (**Fig. 2C-D**). This transition from elastic to elasto-plastic deformation also coincided with the first visible slip activity and the first spike $(dh/dl)$ observed in **Fig. 2E**.

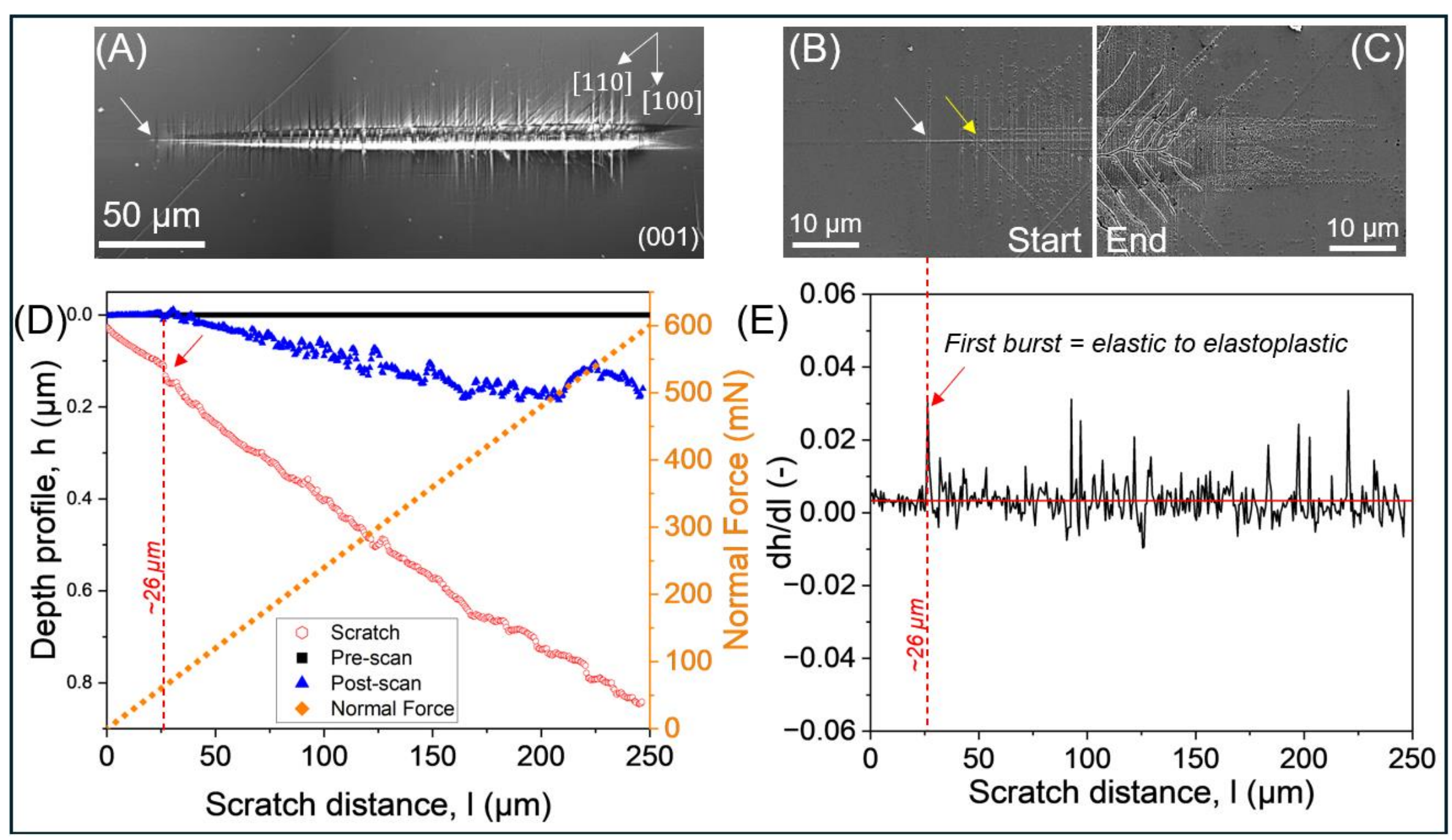

***Figure 2.*** *Microscratching test on the pristine region of single-crystal $SrTiO_3$. (A) Laser scanning microscope image of the microscratch; SEM enlarged views of the start (B) and end (C) of the microscratch, with visible cracks and dislocation etch pits after chemical etching. (D) Plot of surface depth profile against scratch distance of the microscratch, depicting the pre-scan, microscratch, and post-scan. (E) derivative (dh/dl) of profile depth and highlighting the scratch distance corresponding to the first spike. The first spike corresponds to the transition from elastic to plastic deformation during microscratching test, visually evidenced from the SEM images in (B).*

The activated slip traces right after this transition are associated with the $\{110\}_{45}$ slip planes (vertical and horizontal slip lines evidenced by the dislocation etch pits in **Fig. 2B**), in line with the room-temperature active slip systems as previously reported for $SrTiO_3$ [27]. Interestingly, with further increasing load, at a scratch distance of ~50 μm (highlighted with a yellow arrow in **Fig. 2B**), two additional $\{110\}_{90}$ -type slip traces are observed, which are 45° inclined to the scratching direction. Detailed 3D slip traces and dislocation arrangement in $SrTiO_3$ have been presented elsewhere [28]. Continuing with the scratching, at higher loads (above 150 mN, ~75 μm scratch distance), the first signs of median/radial cracks were observed, marking the transition from purely dislocation-mediated plasticity to the onset of cracking. The cracking formation will be discussed in **Sec. 4.2**.

### 3.2. Effect of seeded dislocation density on scratch deformation and cracking behavior

#### 3.2.1 Dislocation density and crack suppression

One of the key reasons for the limited plasticity before cracking in ceramics under mechanical loading is that dislocations and dislocation sources are not readily available. Most conventional ceramics are synthesized at elevated temperatures from powders, where pre-existing dislocations are scarcely present. Compared with metals, much lower densities of dislocations are often reported in single [29] and polycrystalline [30] $SrTiO_3$ (~$10^{10}$ $m^{-2}$). Hence, mechanically seeded dislocations, achieved at a low stress level [31] just sufficient to activate the slip systems at room temperature yet still below the stress threshold for crack formation, have been attempted on various ceramics using Brinell indentation and scratching tests [19, 20, 23].

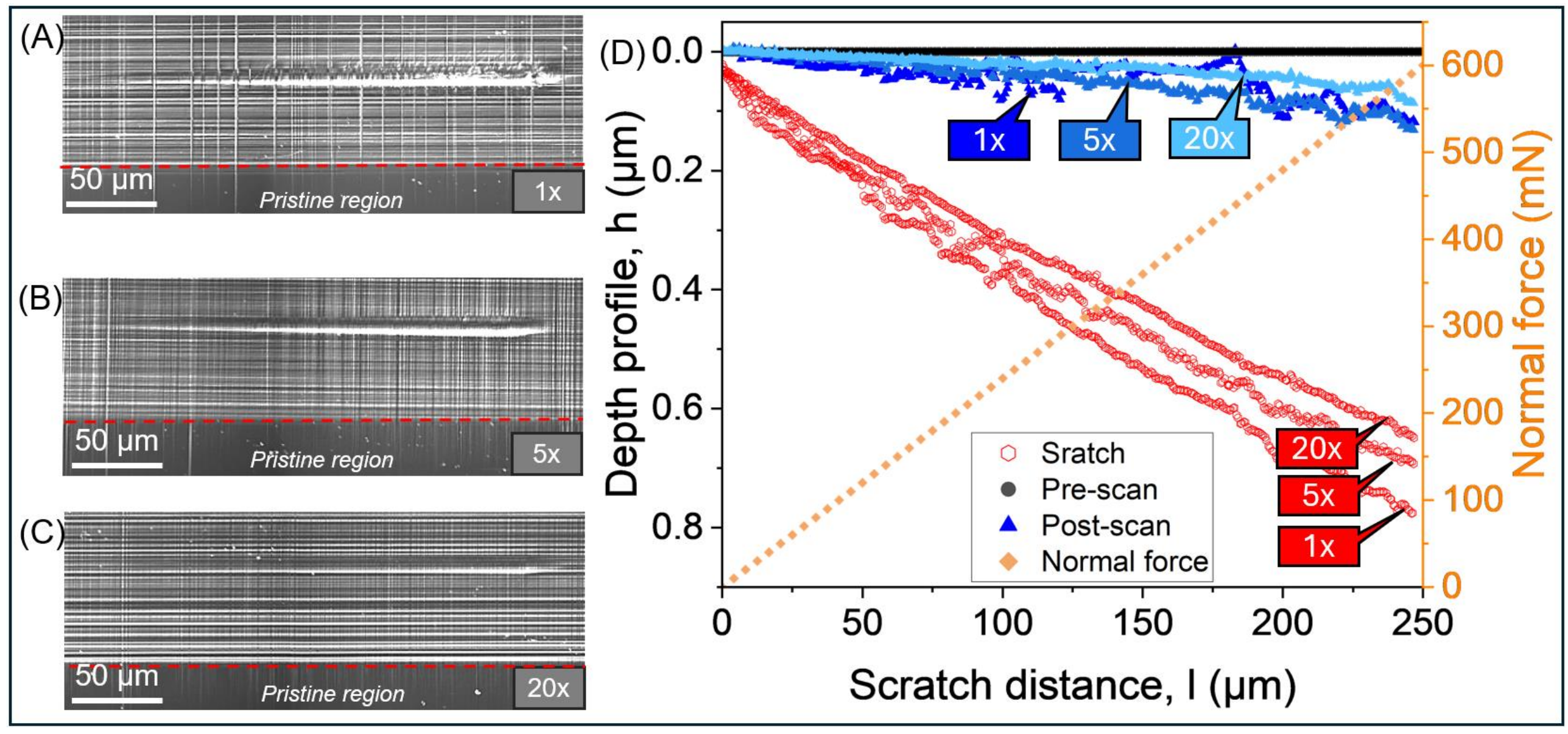

***Figure 3.*** *Laser scanning microscope image of single crystal $SrTiO_3$, depicting the microscratch tracks on pre-inserted Brinell indenter scratch (A) 1x, (B) 5x, and (C) 20x. x denotes the number of Brinell indenter scratch passes. (D) Plot of depth profile against scratch distance of the microscratch, depicting the pre-scan, microscratch, and post-scan, corresponding to (A-C), while the normal force is overlaid.*

In **Figs. 3A-C**, the scratch tracks induced by Brinell indenter cyclic scratching, overlaid with the microscratches, are presented. The red dashed lines indicate the visible boundaries where the slip traces terminate vertically. The dislocation density after 1x, 5x, and 20x Brinell cyclic scratches is of the order of magnitude of ~$10^{11}$ $m^{-2}$, ~$10^{13}$ $m^{-2}$, and ~$10^{14}$ $m^{-2}$, respectively, as previously reported by the current authors using transmission electron microscopy [24]. Unlike the pristine sample subjected to microscratch (**Fig. 2**), these samples with seeded dislocations, regardless of the number of scratch passes, show no clearly defined elastic region and are dominated by plastic deformation (**Fig. 3D**). Nevertheless, distinct differences appear in the tribomechanical response as dislocation density increases. Focusing on the end of the microscratch track, the maximum depth of the microscratch decreases with increasing number of Brinell indenter scratch passes. The maximum depths were measured to be 0.8 µm, 0.65 µm, and 0.6 µm (**Figs. 3D**), respectively. This trend is qualitatively consistent with Taylor hardening [32], where increasing dislocation density raises the resistance to further plastic flow through dislocation-dislocation interactions. A similar trend has been reported in MgO for Knoop hardness testing [10, 33] and in $SrTiO_3$ via Vickers hardness testing [18, 34].

Notably, radial cracking on the {110} planes was also observed as the load increased during microscratching test. In **Fig. 3A**, with a lower seeded dislocation density, the strong contrast within the microscratch track, captured by the laser-scanning microscopy images, suggests the presence of radial cracks. These radial cracks emanate from the microscratch, predominantly at higher loads (i.e., for extended scratch distances above 100 μm and normal force ~250 mN).

### 3.2.2. Change in crack geometry with increasing dislocation density

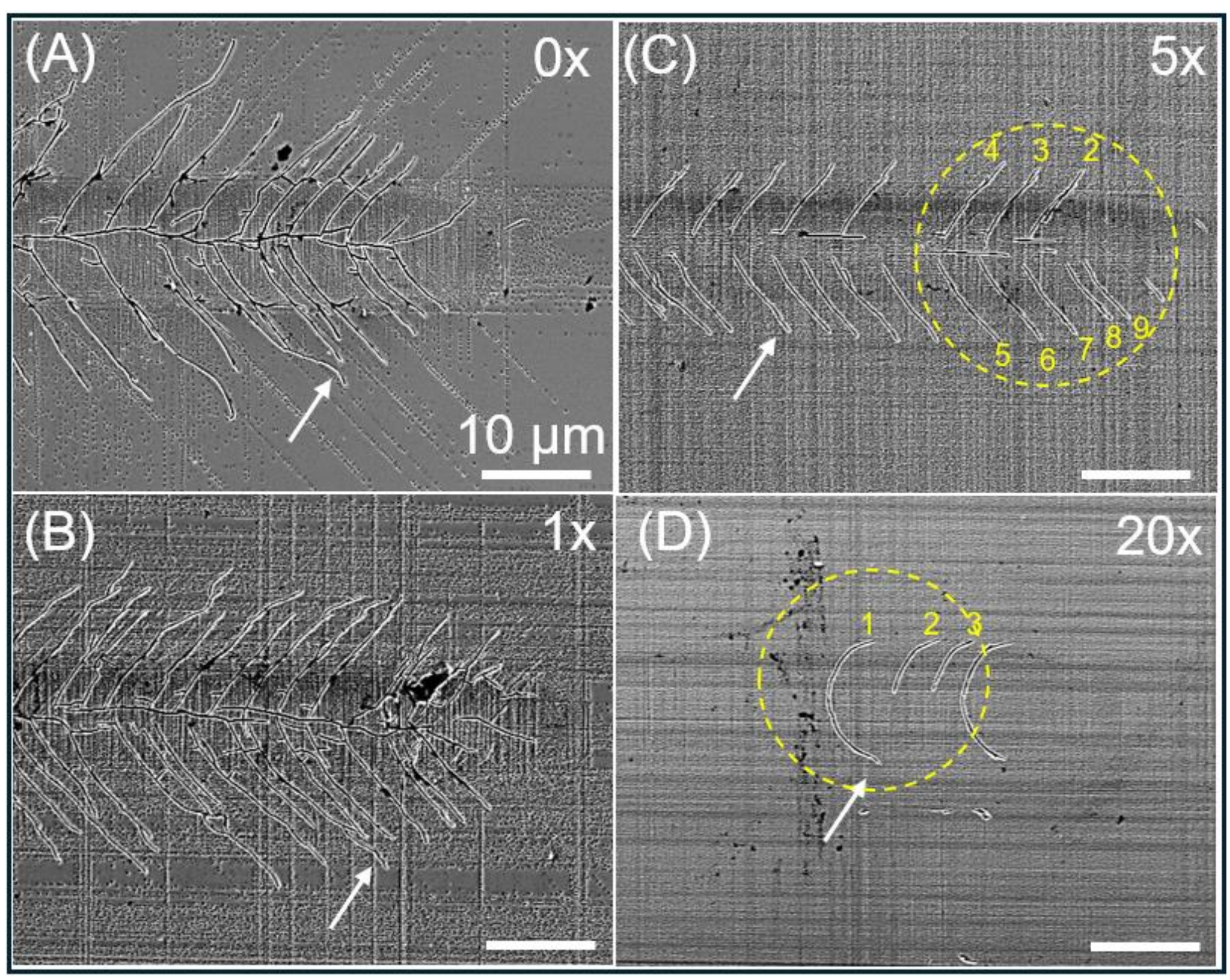

***Figure 4.*** *Impact of dislocation density on crack suppression during microscratching on single-crystal $SrTiO_3$. SEM images of the end of the microscratch: (A) 0x, (B) 1x, (C) 5x, (D) 20x. Yellow dashed circles indicate the intended area for the FIB-milled micropillars in* ***Fig. 6A, B****.*

For detailed local observation, SEM is used to visualize the dislocation and cracking behavior. The SEM imaging was focused more on the high-load region of the microscratching track, as shown in **Fig. 4**. In **Figs. 4A-D**, SEM observations of the end of the microscratch (~600 mN force) for different pre-seeded dislocation densities, including the pristine condition, are presented. In the pristine (0x) and 1x regions, long median/radial cracks were observed along the scratch track (**Fig. 4A, B**). These cracks appear to follow the activated glide planes/dislocation arms as evidenced by the etch pit direction. With increasing Brinell indenter scratch passes (5x, hence a higher dislocation density), the median/radial crack length and crack density decreased markedly (**Fig. 4C**), indicating progressive crack suppression by the pre-seeded dislocations. In **Fig. 4D** (20x, highest dislocation density), the cracks present in the vicinity of the threshold load were predominantly partial cone cracks.

Furthermore, SEM imaging of the 20x regions was performed to identify the threshold load at which partial cone cracks emerge. The microscratch in **Fig. 5A** is overlaid with a schematic of six SEM image partitions taken at increasing loads up to 600 mN, with the areas at 500 mN and 600 mN highlighted. **Figures 5B-C** present SEM images of the 400-500 mN and 500-600 mN load partitions. Consistent with the observations in **Fig. 4D**, the partial cone cracks persist in **Fig. 5C**, with an onset at ~560 mN. These observations show that mechanically seeded dislocations strongly alter the microscratching damage response, delaying crack formation and shifting the dominant crack pattern. This change may arise from the combined influence of dislocation-mediated plasticity, associated residual compressive stress fields, and work hardening. However, the present experiments cannot separate the individual contributions of these effects. A detailed discussion of the change in crack geometry is presented later in Sec. 4.

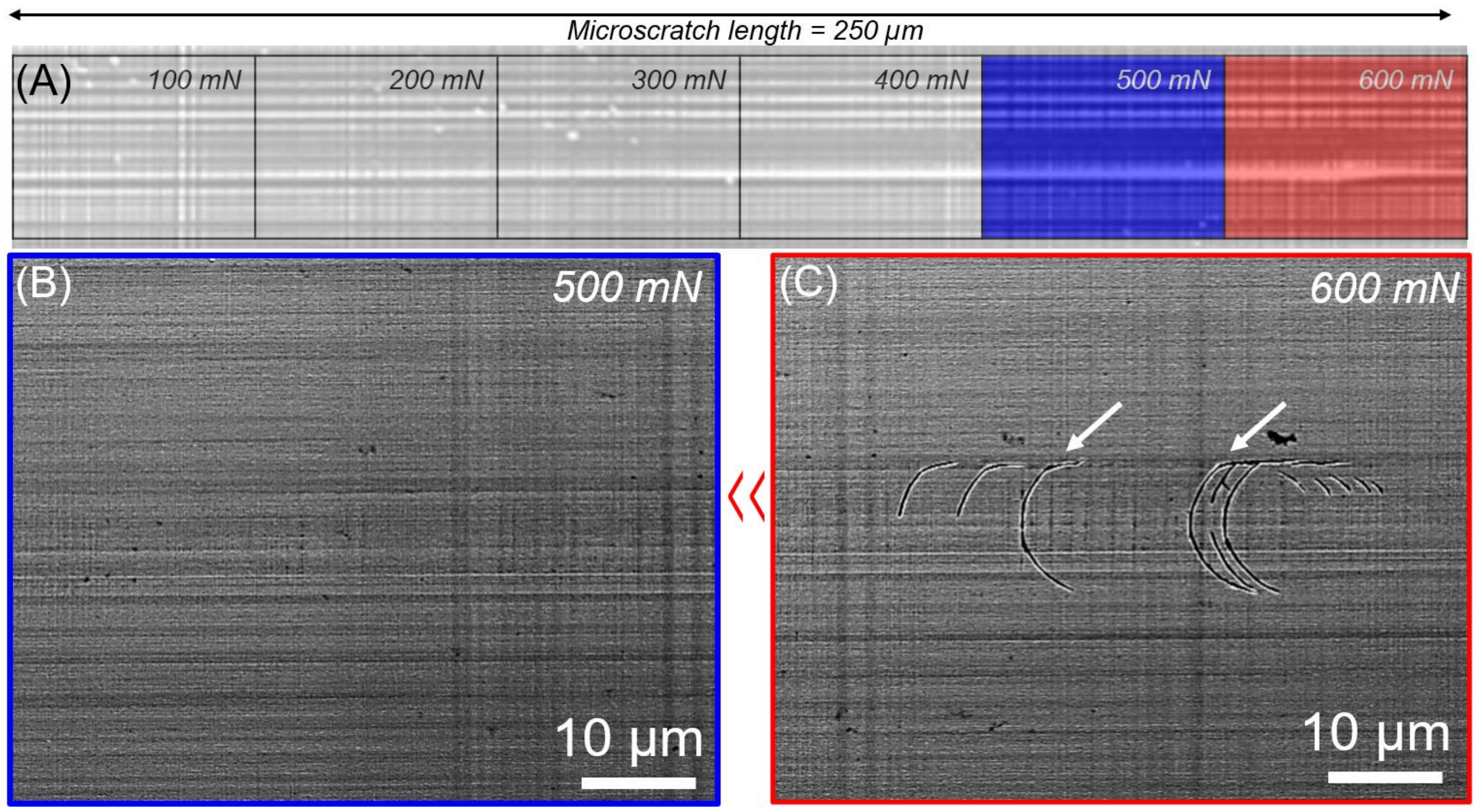

***Figure 5.*** *(A) Schematic of the load partitions (up to 600 mN) overlaid on the microscratch. Two SEM image partitions, corresponding to 500 mN and 600 mN loads, are highlighted. Scanning electron microscopy image of nominally undoped* $SrTiO_3$ *with seeded dislocations after 20x and microscratch forces of (B) 400-500 mN partition, depicting plasticity with high dislocation density and (C) 500-600 mN partition, highlighting the emergence of partial cone cracks. The onset of partial cone cracks is at ~560 mN.*

### 3.2.3. Crack geometry-dependent crack depth

Given that SEM imaging only presents surface information, Nano-CT was employed to quantify the crack depth associated with the two crack geometries (**Fig. 4C-D,** yellow dashed circles) observed after the microscratching test. Representative micropillars of lengths ~20 μm, containing median/radial and partial cone cracks, were reconstructed, and the depth of each visible crack was measured from the top surface to the deepest point in the crack front, as presented in **Fig. 6**. **Figures 6A-B** present the reconstructed volume containing the median/radial and partial cone cracks, respectively. Dashed white curves indicate the cutoff for the cropped segments of the micropillars presented in **Fig. 6A1** and **B1**. The corresponding segmented crack traces are further enhanced with color stripes, and numbers serve as identifiers. Note that the labeling of the crack numbers in **Fig. 6A1** and **B1** corresponds to those in the SEM images in **Fig. 4C-D**. The colored overlay in **Fig. 6A1**and **B1** is used only to aid the visualization of the individual cracks.

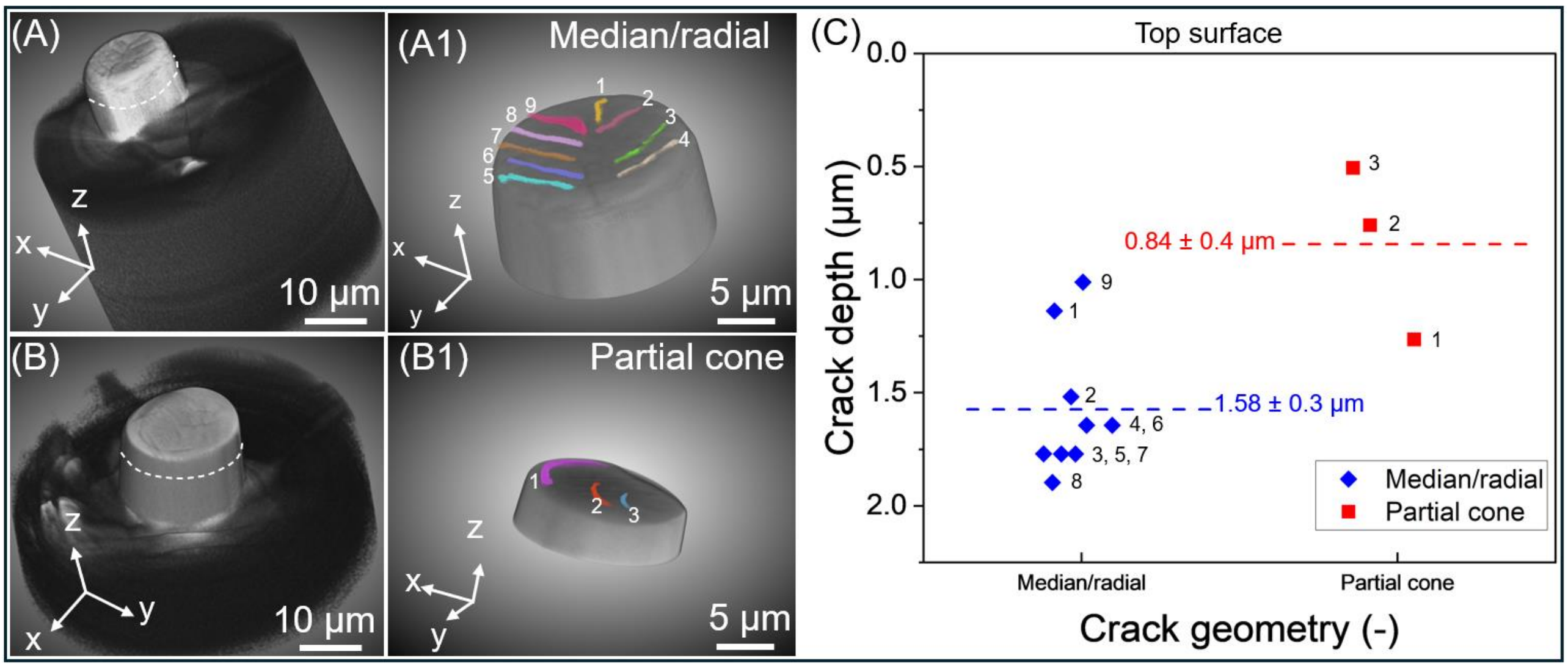

***Figure 6.*** *3D reconstruction of crack geometries after Nano-CT depicting the two crack geometries. Overview of the micropillars for (A) Median/radial crack and (B) Partial cone crack geometry. Dashed white markers represent the cutoff of the cropped section in (A1) and (B1). Colored overlays in A1 and B1 are only for easy visualization of the cracks. (C) Variations in the crack depth with respect to crack geometry. Average crack depths of 1.58 ± 0.3 μm and 0.84 ± 0.4 μm for the median/radial and partial cone cracks, respectively. The numbers correspond to individual crack positions in (A1) and (B1).*

The quantified crack depths are summarized in **Fig. 6C**. The top surface is defined as the reference position (0 μm), and increasing depth (with positive values) is plotted downwards following the procedure described in **Sec. 2.5**. The median/radial crack depths are between 1 and 1.9 μm, with an average depth of 1.6 ± 0.3 μm. In contrast, the partial cone crack shows depths between 0.5 and 1.3 μm, with an average depth of 0.8 ± 0.4 μm. Hence, within the measured crack population, the median/radial cracks are approximately twice as deep as the partial cone cracks. Meanwhile, crack evolution during laser/FIB milling cannot be fully excluded. Material removal around the region of interest may relax or redistribute residual stresses associated with the scratch damage field [35]. Notably, crack position 1 in **Fig. 6A1** appears to have been introduced during laser/FIB milling, as it was not observed in the SEM images in **Fig. 4C**. This crack appears much shorter compared to the 'native' cracks prior to milling.

### 3.3. Friction and surface topography

So far, the influence of dislocation density on the elastic, elastoplastic, and cracking behavior of single-crystal $SrTiO_3$ via the two-step scratching approach was presented. Here, we present the possible impact of surface topography (before and after Brinell indenter scratching and the number of scratch passes) on the coefficient of friction (COF). The variation in the COF as a function of the scratch distance and the dislocation densities is presented in **Fig. 7A**. As expected, there is a slight increase in the COF with the scratch distance, but overall, the COF for all test conditions has a very low value (below 0.08). At scratch distances greater than 100 µm, cracking becomes more obvious, which influences the COF. Interestingly, an overall comparison of the peak COF in **Fig. 7B** shows first an increase and subsequently a decrease in the COF with increasing Brinell indenter scratch passes (i.e., increased dislocation densities). This trend may not be surprising, as the COF also correlates with the surface roughness. The increase in surface roughness after 1x scratch pass and the subsequent decrease could be attributed to the discrete distribution of slip traces (formation of slip steps). With an increasing number of Brinell passes up to 20x, slip traces are more uniformly distributed, leading to decreased surface roughness and, in turn, reduced COF.

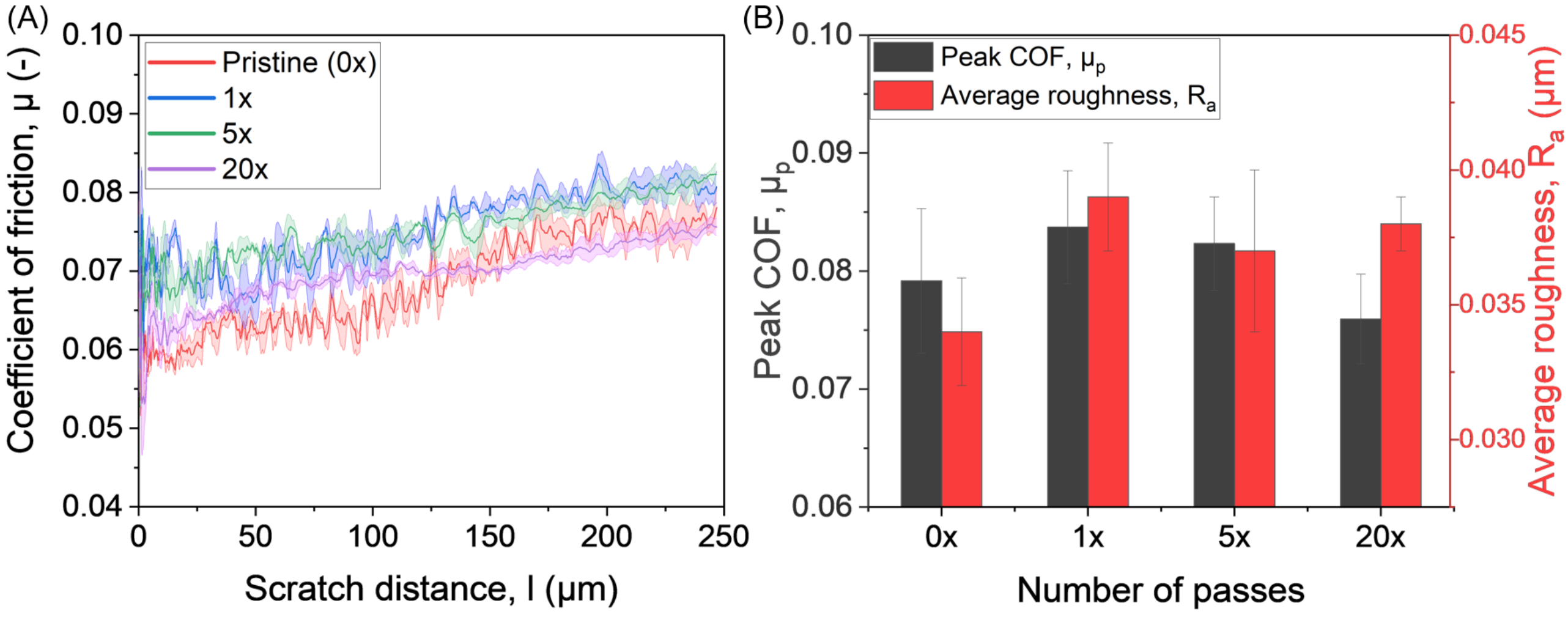

***Figure 7.*** *(A) Coefficient of friction (COF) plotted against scratch distance. The shaded area represents the standard error. (B) Peak COF,* $\mu_p$*, and average surface roughness* $R_a$ *with respect to the number of Brinell indenter scratch passes. Dashed lines show the trend of peak COF and average surface roughness* $R_a$ *with the number of Brinell indenter scratch passes.*

## 4. Discussion

### 4.1. Change in crack geometry with respect to dislocation density

#### 4.1.1 The median/radial crack model

The results in **Sec. 3** demonstrate that mechanically seeded dislocations strongly influence the tribomechanical response of single-crystal $SrTiO_3$. The cracking behavior in the pristine region of $SrTiO_3$ shares common features with brittle ceramics, exhibiting limited plasticity before the onset of cracks and followed by predominant median/radial cracks. The formation of median/radial cracks can be described using the elastic/plastic indentation framework established by Lawn, Evans, and Marshall (the LEM model) [36]. In this model, the indentation stress field is separated into an elastic component and a residual component associated with the plastic zone beneath the contact. For a well-developed median/radial crack, the driving force scales as follows [36, 37]:

$$K_I = \chi \frac{P}{c^{1.5}} \tag{1}$$

where $K_I$ is the mode *I* stress intensity factor, $P$ is the applied normal load, $c$ is the median/radial crack length, and $\chi$ is a coefficient that depends on the indenter/material. Cracking occurs when:

$$K_I \geq K_{IC} \tag{2}$$

where $K_{IC}$ is the fracture toughness. For indentation-induced median/radial cracks, it is more commonly described as follows [36]:

$$K_r = \xi (\frac{E}{H})^{0.5} \frac{P_c^*}{c^{1.5}} \tag{3}$$

where $K_r$ is the stress intensity factor for the median/radial cracks, with $E$ and $H$ being the elastic modulus and the hardness, respectively. $\xi$ is a geometry-dependent constant, and $P_c^*$ is the critical load leading to unstable crack propagation. Although the LEM model was originally proposed for point-load indentation, the contribution of the tangential force during sliding contact is accounted for by $P_c^* = P_c f$ [38], where $f$ is the COF measured during the scratch test (**Fig. 7A**). This relationship is useful for interpreting the current results. If the values for the applied load and fracture toughness are fixed, an increase in apparent hardness reduces the median/radial crack length according to Eq. (1). Therefore, the observed decrease in crack length and crack density with increasing Brinell indenter scratch passes (hence increased hardness) is consistent with a reduction in the median/radial crack-driving tendency. To be more specific, the increase in apparent hardness with dislocation density can be rationalized, in a simplified manner, using the Taylor hardening criterion [32, 39]:

$$\sigma_y = \sigma_0 + \alpha G b \sqrt{\rho} \tag{4}$$

With $\sigma_y$ being the flow stress, $\sigma_0$ the lattice friction stress, $\alpha$ a constant, $G$ the shear modulus, $b$ the Burgers vector, and $\rho$ the dislocation density. Since hardness scales approximately with the flow stress $H \approx C\sigma_y$ (*C* being the constraint factor), hence, the hardness can be expressed as follows:

$$H(\rho) \approx H_0 + C\alpha G b \sqrt{\rho} \tag{5}$$

Here, $H_0$ is benchmarked to be the hardness for the 1x Brinell scratch pass, where the dislocation density is relatively low to contribute little to work hardening but sufficiently high to circumvent dislocation nucleation. Following Eq. (5), increasing the dislocation density raises the resistance to further plastic flow and reduces the penetration depth during microscratching, as has been reported previously during point-load indentation [18, 23, 34]. This explains why the high-dislocation-density regions show lower scratch depth under the same applied load. With a combination of Eqs. (3), (4), and (5), the crack length $c$ scales with the hardness (i.e., dislocation density). Hence, the mechanically seeded dislocations have two connected effects: widening the plasticity window, and subsequent work-hardening at high dislocation densities.

### 4.1.2 The partial cone crack model

Here, the Lawn-Wiederhorn-Roberts (LWR) model is more appropriate [40] for the partial-cone crack configuration. The LWR model was originally proposed to model cone cracking during spherical indentation in brittle materials such as glass. During sliding contact, tensile stress peaks locally at the rear of the contact circle, where the crack initiates and forms a partial-cone crack. Since the fracture originates from the surface, the partial cone crack appears shallower than the median/radial cracks (see **Fig. 6C**).

The LWR model can be extended to treat the sliding contact by superposing a tangential load $fP$, onto the normal load $P$, where $f$ is the COF. In this formulation, the effective load [40] becomes:

$$P' = P(1 - f^2)^{0.5} \tag{6}$$

correspondingly, combining Eq. (1) into Eq. (6) leads to:

$$K_p = \eta \frac{P'}{c^{1.5}} \tag{7}$$

Where $K_p$ is the stress intensity factor for the partial-cone crack, $\eta$ is a dimensionless geometric constant, and other parameters are as previously described. This relation directly shows that sliding friction contributes to the crack-driving condition through COF. However, $K_p$ is weakly dependent on the COF, especially when the COF value is small as in the current case. The stress intensity factor as a function of the load at which a particular crack geometry dominates over the dislocation density, using the modified Eqs. (3) and (7) is described as follows:

$$K_{r,LEM} = \xi (\frac{E}{H})^{0.5} \frac{P}{c^{1.5}} \tag{8a}$$

$$K_{p,LWR} = \frac{\eta P(1-f^2)^{0.5}}{c^{1.5}} \tag{8b}$$

This point is especially relevant to the present results because the measured COF from the experiment (**Fig. 7**) remains within a relatively narrow range of 0.06-0.08 (therefore, $f^2 \to 0$). Hence, the friction-controlled LWR contribution to partial cone cracking is considered minor as the pre-seeded dislocation density increases.

This provides a useful interpretation for the schematic representation in **Fig. 8B**. At low dislocation density, the stress intensity factor for the median/radial is higher than that of the partial cone crack $K_r > K_p$, and the median/radial cracks dominate the observed damage. With increasing dislocation density, the apparent hardness increases while the COF remains nearly unchanged, so the LWR-type friction-controlled contribution remains almost constant. At the transition point ($K_r \approx K_p$), in **Fig. 8B**, the LEM-type median/radial cracking is significantly suppressed. Then the crack initiation by the sliding force becomes more favorable with $K_r < K_p$, and the dominant visible crack mode changes to partial cone cracking at high load/dislocation density (when the local tensile stress is sufficiently high). Additionally, the Nano-CT analysis provides a subsurface validation of the crack geometries inferred from SEM. An almost twofold difference in crack depths between median/radial and partial-cone cracks was observed. Although complete crack suppression was not achieved beyond 560 mN load, the dominant partial cone cracks are near-surface rather than median/radial cracks (**Fig. 6C**).

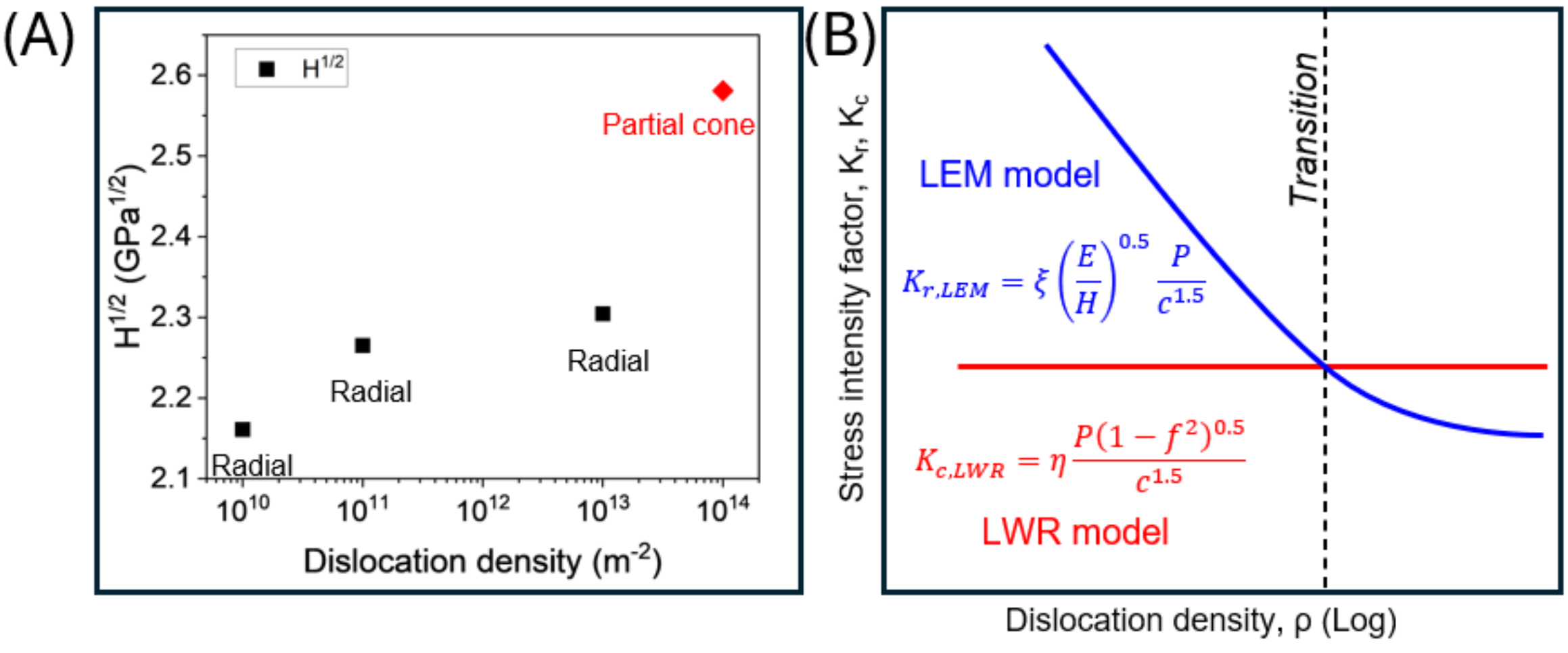

***Figure 8.*** *Dislocation-density-dependent plasticity and cracking response during microscratching on single-crystal $SrTiO_3$. (A) Variation in the square root of the hardness with respect to dislocation density (log) is plotted at a fixed threshold load of 560 mN, ensuring cracking under all conditions. (B) Schematic of the stress intensity factor as a function of dislocation density, showing the transition from radial crack-dominated (Lawn-Evans-Marshal, LEM model) [36] to the partial cone crack-dominated (Lawn-Wiederhorn-Roberts, LWR model) [40]. Note: The logarithmic plot preserves the relative spacing, which spans several orders of magnitude and is consistent with the expected power-law dependence of hardness in Eq. (5).*

In addition to dislocation-mediated resistance to damage (crack initiation and propagation), residual compressive stresses introduced during mechanical dislocation seeding also contribute to the observed crack suppression and shortening. Previous studies on plastically deformed $SrTiO_3$ and MgO have shown that mechanically seeded dislocations are accompanied by residual compressive stress fields which reduce the effective tensile driving force for crack opening and thereby limit crack propagation [18, 23, 24, 34]. Such stresses are expected to be particularly relevant for the median/radial cracks, whose propagation is controlled by tensile opening stresses beneath the scratch track. Preuß et al. [23] quantified and decoupled the influence of residual compressive stress via heat treatment in MgO and still found a significant crack shortening attributed to mechanically seeded dislocations. However, the present study did not measure residual compressive stresses directly; therefore, their contribution cannot be quantitatively separated from the effects of increased apparent hardness, elastic recovery, and crack suppression. The present results should therefore be interpreted as a combined outcome of dislocation-mediated plasticity and possible residual compressive stress fields.

**Figure 9** summarizes the impact of seeded dislocations on the plasticity and cracking behavior of single-crystal $SrTiO_3$ within the *deformation map*. At low loads (less than 75 mN), the pristine condition first shows a small elastic window, followed by plasticity without visible cracking. A window of plasticity is observed that widens with increasing dislocation density, consistent with the findings from previous studies via point-load indentation on single-crystal $SrTiO_3$ [24, 34]. However, at low dislocation densities (pristine sample), there is an onset of median/radial cracking at load ~ 125 mN. This crack geometry is consistent with the elasto-plastic crack driving force described by the LEM model [36]. With increasing dislocation density, the onset of cracking was further delayed until a threshold load of 560 mN, at which partial cone cracking emerges. These threshold values were experimentally determined from the first visible dislocation etch pit (plasticity) and the cracking load from post-analysis after microscatching. The high-load/high dislocation density controlled crack geometry is consistent with the LWR sliding-contact model [40]. This increases the dislocation density, delays cracking, and reduces the severity of median/radial crack penetration until the transition from residual stress-driven median/radial cracks to the sliding-contact-induced partial-cone cracks.

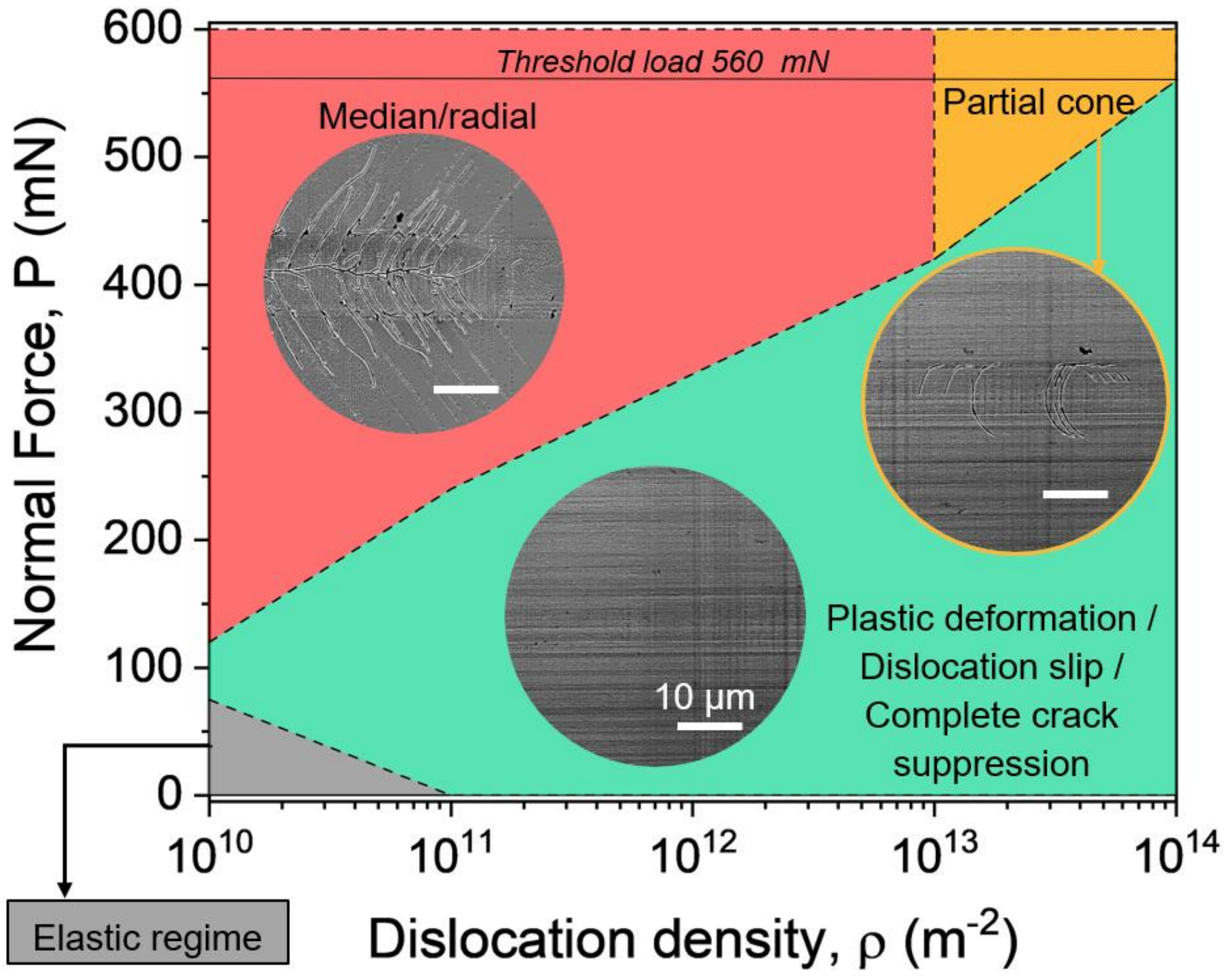

***Figure 9.*** *Deformation map for the microscratching test on single-crystal $SrTiO_3$ using a nominal tip radius of 30 µm. The figure depicts dislocation-density-dependent plasticity and cracking responses. In the pristine sample, at low loads, elastic deformation occurs, followed by an increasing plasticity window with dislocation density. At the upper threshold load, cracking dominates. The vertical boundary indicates the transition from median/radial crack to partial cone crack mode. Dashed lines indicate the transition, with the inclined boundary representing the experimentally observed onset of cracking that separates the elastoplastic deformation regime from the cracking regime.*

### 4.2. Other contributing factors, COF, and surface roughness

In addition to dislocation density and residual stress, surface topography and friction must be considered because cyclic Brinell ball scratching changes the initial surface roughness before microscratching. The COF increases with microscratch distance/load for all conditions (**Fig. 7A**), as expected in a load-ramped scratch test because the contact pressure, penetration depth, and damage accumulation increase continuously along the scratch path. At larger scratch distances/loads (>100 µm / 300 mN), where cracking becomes more pronounced, the COF is also expected to be affected by crack opening and the increased contribution of plowing acting as resistance to the scratching.

However, the overall variation in COF is relatively small, specifically 0.06-0.08 between the pristine region and regions with pre-seeded dislocations via Brinell indenter scratching. Similarly, the variations in surface roughness after Brinell indenter scratching do not scale directly with the observed transition from median/radial cracks to partial cone cracks. Although the 1x Brinell indenter scratched region shows an increase in roughness, likely due to discrete slip-step formation (**Fig. 3A**), the roughness decreases again with increasing number of passes as slip traces become much finer and more uniformly distributed. This trend does not explain the monotonic suppression of median/radial cracking with increasing dislocation density.

These observations suggest that friction and surface topography are not the primary factors controlling the crack suppression observed in this work. Instead, they may act as secondary contributions to the scratch response. The relatively small change in COF supports our interpretation that mechanically seeded dislocations mainly govern the crack-mode transition, providing carriers for plastic deformation and work hardening (**Fig. 8A**) and resulting in crack shortening.

## 5. Conclusions

Here, we demonstrate that mechanically seeded dislocations accompanied by residual compressive stress can effectively modify the microscale tribological response of nominally undoped single-crystal $SrTiO_3$ at room temperature by first seeding dislocations into the crystals via Brinell indenter scratching (2.5 mm tip diameter) and subsequently performing load-ramped microscratching tests using a 30 μm (nominal tip radius) spherical indenter. The main findings are summarized as follows:

(1) In the pristine region with a low dislocation density of ~$10^{10}$ $m^{-2}$, microscratching-induced deformation proceeds through a clear elastic-to-elastoplastic transition, followed by median/radial cracking at higher loads. In contrast, the dislocation-rich region exhibits immediate plastic accommodation, reduced scratch penetration depth, and progressive suppression of median/radial cracks as dislocation density increases, accompanied by an increase in residual compressive stress. At the highest dislocation density of ~$10^{14}$ $m^{-2}$, cracking is suppressed up to ~560 mN, with partial cone cracking emerging afterward.

(2) The transition from median/radial cracking to partial cone cracking indicates that dislocation engineering not only delays damage towards higher loads but also changes the dominant crack-driving tendency during microscratching. The relatively small variation in the coefficient of friction and surface roughness further supports the conclusion that the observed crack suppression is primarily controlled by increased dislocation density, residual compressive stresses, and the accompanying work hardening induced during the dislocation-seeding process.

(3) A dislocation-density-based deformation map in its preliminary form is constructed based on the load and dislocation density for the given sliding indenter tip radius. In this regard, pre-seeded dislocations can widen the damage-free range and open an application window for future functional application scenarios with higher demand for damage tolerance.

These findings show that dislocation engineering can modify the near-surface damage tolerance of plastically deformable ceramics. Future work will examine the tribomechanical response using spherical tips of different sizes to assess the size effect on the competition between plasticity and cracking, as well as the impact of defect chemistry (e.g., point defects). Varying the tip radius will change the contact stress field, contact area, and plastic zone size, providing a route to clarify how dislocation density, defect chemistry, and contact geometry jointly and independently control the mechanical deformation of ceramics subjected to sliding contact.

**Acknowledgement**

C. Okafor acknowledges the financial support from the Deutsche Forschungsgemeinschaft (DFG, grant No 510801687). X. Fang and O. Preuß are also supported by the European Union (ERC Starting Grant, Project MERCERDIS, grant No. 101076167). Views and opinions expressed are, however, those of the authors only and do not necessarily reflect those of the European Union or European Research Council. Neither the European Union nor the granting authority can be held responsible for them. YG acknowledges the support from the US National Science Foundation (CMMI 2412544). KP and CK acknowledge funding from the DFG within the Collaborative Research Center (TRR 188, "Damage Controlled Forming Processes", 278868966) in project B03 "Understanding the damage initiation at microstructural scale". This work was partly carried out with the support of the Karlsruhe Nano Micro Facility (KNMFi, www.knmf.kit.edu), an Open Access Research Infrastructure within the Karlsruhe High Technology Hub at the Karlsruhe Institute of Technology (KIT – The University in the Helmholtz Association, www.kit.edu).

**Author contributions:**

Conceptualization: XF; Methodology: CO, TC, UB, DE, XF; Investigation: CO, OP, TC, UB, KP, YG, CK, XF; Visualization: CO, OP, TC, KP; Formal analysis: CO, OP, TC, KP; Funding acquisition: CK, XF; Resources: TC, XF; Project administration: XF; Supervision: XF; Writing – original draft: CO; Writing – review and editing: CO, OP, TC, UB, DE, KP, YG, CK, XF.

**Conflict of interest:** The authors declare no known competing interests.

**Data and materials availability:** All data are available in the main text.

**Disclosure of use of AI tools:** No AI-generated text, code, figure, or analysis was included in the manuscript. The authors wrote, reviewed, and approved all content.